\documentstyle[preprint,eqsecnum,aps,prd]{revtex}
\def\a{\alpha}
\def\ap{{\alpha '}}
\def\nabboth{\tensor{\nabla}}
\def\parboth{\tensor{\partial}}
\def\Ca{\langle x''||C_\a||x'\rangle}
\def\Cu{\langle x''||C_u||x'\rangle}
\def\rb{{\overline r}}
\def\rnb{{\overline r(n)}}
\def\rpmnb{{\overline r(\pm n)}}
\def\Cr{\langle x''||C_r||x'\rangle}
\def\Cl{\langle x''||C_l||x'\rangle}
\def\Cb{\langle x''||C_b||x'\rangle}
\def\Crb{\langle x''||C_\rb||x'\rangle}
\def\CrbN{\langle x''N||C_\rb||x'0\rangle}
\def\Crnb{\langle x''||C_\rnb||x'\rangle}
\def\CrnbN{\langle x''N||C_\rnb||x'0\rangle}
\def\D{{\cal D}}
\def\op{\omega_{\hbox{\tiny\bf p}}}
\def\ok{\omega_{\hbox{\tiny\bf k}}}
\def\x{\hbox{\scriptsize\bf x}}
\def\xo{\hbox{\scriptsize\bf x}_0}
\def\exi{\hbox{\scriptsize\bf x}_i}
\def\p{\hbox{\scriptsize\bf p}}
\def\k{\hbox{\scriptsize\bf k}}
\def\kp{\hbox{\scriptsize\bf k}_\perp}
\def\ko{\hbox{\scriptsize\bf k}_0}
\def\kop{\hbox{\scriptsize\bf k}_{0\perp}}
\def\kc{\hbox{\scriptsize\bf k}_c}
\def\lo{{\lambda_1}}
\def\lt{{\lambda_2}}
\begin{document}
\draft
\preprint{submitted to {\it Physical Review D}}
\title{Spacetime alternatives in relativistic particle motion}
\author{John T.~Whelan\thanks{Electronic Mail:
whelan@sbphy.physics.ucsb.edu}}
\address{Department of Physics, University of California,
Santa Barbara, California 93106-9530\\
and
Isaac Newton Institute for Mathematical Sciences, 20 Clarkson Road,
Cambridge, CB3 0EH, United Kingdom}
\date{draft June 9, 1994}
\maketitle
\begin{abstract}
Hartle's generalized quantum mechanics formalism is used to examine
spacetime coarse grainings, i.e., sets of alternatives defined with
respect to a region extended in time as well as space, in the quantum
mechanics of a free relativistic particle.  For a simple coarse
graining and suitable initial conditions, tractable formulas are found
for branch wave functions.  Despite the nonlocality of the
positive-definite version of the Klein-Gordon inner product, which
means that nonoverlapping branches are not sufficient to imply
decoherence, some initial conditions are found to give decoherence and
allow the consistent assignment of probabilities.
\end{abstract}
\narrowtext
\section{Introduction}

     One of the reasons why we expect a standard quantum mechanics,
described by states on a spacelike surface, to be inadequate to
describe quantum gravity is that the notion of ``spacelike'' should be
ill-defined in a theory where the metric itself is behaving quantum
mechanically.  Standard quantum mechanics makes reference to spacelike
surfaces not only in its description of the state of the system ``at a
moment of time'', but also in the very alternatives for which it makes
predictions.  A theory which predicts spacetime probabilities, such as
the probability that a particle passes through an extended region of
spacetime during its trajectory, can thus be thought of as one step on
the road towards a quantum theory of gravity.  Spacetime alternatives
in nonrelativistic quantum mechanics have been considered in the past
by Feynman\cite{feyn}, Yamada and Takagi
\cite{Y&T}, and Hartle\cite{nonrel}.  

     The present work considers spacetime alternatives for the quantum
mechanics of a free relativistic particle.  This is not meant as a
quantum theory of actual relativistic particles (which are described
by quantum field theory) but rather as a toy model for quantum
cosmology.  The relativistic particle is a better analogy to gravity
than is the nonrelativistic particle, because it exhibits a single
reparametrization invariance, which can be though of as a subset of
the diffeomorphism invariance exhibited by general relativity.

     We will use Hartle's generalized quantum mechanics
formalism\cite{gqm}.  The three fundamental elements of this theory
are the possible histories of the system (``fine-grained histories''),
allowable partitions of the histories into classes $\{c_\a\}$ so that
each history is contained in exactly one class (``coarse grainings''),
and a complex matrix $D(\a,\ap)$ corresponding to each coarse graining
(``decoherence functional'').  The decoherence functional must satisfy
the following properties:
\begin{mathletters}
\begin{itemize}
     \item ``Hermiticity'':
          \begin{equation}
          D(\ap,\a)=D(\a,\ap)^*;
          \end{equation}
     \item positivity of diagonal elements:
          \begin{equation}\label{pos}
          D(\a,\a)\ge 0;
          \end{equation}
     \item normalization:
          \begin{equation}
          \sum_\a\sum_\ap D(\a,\ap)=1;
          \end{equation}
     \item superposition: If $\{c_\beta\}$ is a coarse graining
constructed by combining classes in $\{c_\a\}$ to form larger
classes (``coarser graining''), i.e., $c_\beta =
\bigcup\limits_{\a\in\beta} c_\a$, the decoherence functional for
$\{c_\beta\}$ can be constructed from the one for $\{c_\a\}$ by
          \begin{equation}\label{super}
          D(\beta,\beta')=\sum_{\a\in\beta}\sum_{\ap\in\beta'}
D(\a,\ap).
          \end{equation}
\end{itemize}
\end{mathletters}

     When the decoherence functional is diagonal, or nearly so:
\begin{equation}
D(\a,\ap)\approx\delta_{\a\ap}p_\a
\end{equation}
[``(medium) decoherence''], then the diagonal elements $\{p_\a\}$
are the probabilities of the alternatives $\{c_\a\}$, and obey
classical probability sum rules.  When the
alternatives do not decohere, quantum mechanical interference
prevents the theory from assigning probabilities to them.

     In this paper, we calculate the decoherence functionals for
certain simple coarse grainings.  Depending on the initial
conditions, some of these sets of alternatives will decohere and
others will not.  The cases which exhibit decoherence provide
predictions for spacetime alternatives, however contrived.

\section{Decoherence functional and class operators}
\subsection{General prescription}\label{setup}

     In constructing a generalized quantum mechanics of the a
free relativistic particle, we follow the procedure described in
\cite{leshouches} (to which the reader is referred for more details).
The general sum-over-histories recipe is this: for a given
class of paths $c_\a$, a class operator $\Ca$ is constructed via
a sum of $\exp (i \hbox{ action})$ over those histories which
start at co\"{o}rdinate point $x'$ (here a point in spacetime),
end at another (spacetime) point
$x''$ and are in the class $c_\a$:
\begin{equation}
\Ca=\sum_{x'\a x''}e^{iS[\text{history}]}.
\end{equation}
The initial and final conditions are expressed in terms of weights (or
``probabilities'') $\{p'_j\}$ and $\{p''_i\}$, respectively, and wave
functions $\{\psi_j(x')\}$ and $\{\varphi_i(x'')\}$, respectively.
Although we do not presuppose the existence of a Hilbert space of wave
functions, it is illustrative to think of the conditions as being
described by initial and final ``density matrices''
\begin{mathletters}
\begin{equation}
\rho'(x'_1,x'_2)=\sum_j\psi_j(x'_1)p'_j\psi^*_j(x'_2)
\end{equation}
and
\begin{equation}
\rho''(x''_1,x''_2)=\sum_j\varphi_i(x''_1)p''_i\varphi^*_i(x''_2)
,
\end{equation}
\end{mathletters}
respectively.  The initial and final wave functions are attached using
a Hermitian\footnote{By which we mean
$\varphi\circ\psi=(\psi\circ\varphi)^*$.} but not necessarily positive
definite inner product $\circ$:
\begin{equation}
\langle\varphi_i|C_\a|\psi_j\rangle
=\varphi_i(x'')\circ\Ca\circ\psi_j(x');
\end{equation}
and finally the decoherence functional is defined as
\begin{equation}\label{dec}
D(\a,\ap)={\sum\limits_{i,j}
p_i''\langle\varphi_i|C_\a|\psi_j\rangle
\langle\varphi_i|C_\ap|\psi_j\rangle^* p_j' \over
\sum\limits_{i,j}p_i''|\langle\varphi_i|C_u|\psi_j\rangle|^2
p_j'},
\end{equation}
where $C_u$ is the class operator corresponding to the class $c_u$ of
all paths.  This construction satisfies all the usual requirements for
a decoherence functional with positivity of diagonal elements
(\ref{pos}) holding as long as the weights $\{p'_j\}$ and $\{p''_i\}$
are non-negative.  Note that the inner product $\circ$ need not be
positive definite to ensure positivity of the decoherence functional.
The superposition property (\ref{super}) holds because the class
operators are constructed linearly, and thus satisfy their own
superposition property:
\begin{mathletters}\label{classsuper}
\begin{eqnarray}
\langle x''||C_\beta||x'\rangle=\sum_{\a\in\beta}\Ca\\
\label{classone}\sum_\a\Ca=\Cu.
\end{eqnarray}
\end{mathletters}

     In our chosen realization of the free relativistic particle in
$(D+1)$-dimensional Minkowski spacetime, the canonical action is
written in the form
\begin{equation}
S_{CAN}=\int_0^1 d\lambda\left(p\cdot{dx\over
d\lambda}-N\,{p^2+m^2\over 2m}\right)
\end{equation}
[where $p^2=p\cdot p=p^\mu p_\mu=-(p^0)^2+{\bf p}^2$] and the
fine-grained histories we sum over are parametrized paths
$\left\{p(\lambda),x(\lambda)\right\}$ through phase space and
multiplier histories $N(\lambda)$.  The multiplier $N$ is a quantity
which classically (i.e., for the path of least action) defines the
relationship between proper time and the arbitrary parameter
$\lambda$: $N={d\tau\over d\lambda}$.  Note that the paths are allowed
to move forward and backward in the ``time'' co\"{o}rdinate $x^0$.
This set of fine-grained histories is Lorentz invariant, as opposed to
a theory which restricts the paths to move forward in time in a given
Lorentz frame.

     Note also that the action is invariant under reparametrizations
of the parameter $\lambda$, if $N$ transforms as the derivative of an
invariant quantity.  Since we only consider
reparametrization-invariant coarse grainings as being physically
meaningful, we may restrict our sum over histories to those histories
which satisfy the ``gauge condition'' ${dN\over d\lambda}=0$.  In this
gauge, we need only integrate over a single $N$, which is the total
proper time of the path.  The theory will turn out to have a closer
correspondence to field theory if we integrate only over positive
values of $N$.  The class operator is thus defined by
\widetext
\begin{equation}\label{class}
\Ca=\int\nolimits_0^\infty dN\int\limits_{x'\a x''}\D x\,\D p
\exp\left[i\int\nolimits_0^N d\tau \left(p\cdot{dx\over
d\tau}-{p^2+m^2\over 2m}\right)\right].
\end{equation}
\narrowtext
[We only wish to consider coarse grainings which restrict the
configuration space path $x(\lambda)$, but it is useful to
express the sum over histories in terms of phase space histories
because the measure for the path integral is then naturally
defined.]

     To specify the inner product $\circ$ we define an
``initial'' spacelike $D$-surface $\sigma '$ and a ``final''
spacelike $D$-surface $\sigma ''$ to the future of the initial one,
and apply the Klein-Gordon inner product on those surfaces:
\begin{mathletters}\label{inner}
\begin{equation}\label{inner'}
\varphi(x')\circ\psi(x')=\int_{\sigma '} d^D\!\Sigma
'{}^\nu\varphi_i^*(x') i\nabboth{}'_\nu\psi_j(x')\end{equation}
and
\begin{equation}\label{inner''}
\varphi(x'')\circ\psi(x'')=\int_{\sigma ''} d^D\!\Sigma
''{}^\mu\varphi_i^*(x'') i\nabboth{}''_\mu\psi_j(x'').
\end{equation}
\end{mathletters}
(Here $\nabboth$ is the usual bidirectional derivative:
$\varphi\nabboth_\mu\psi=\varphi\nabla_\mu\psi-\psi\nabla_\mu\varphi$.)
Thus\footnote{We have, of course, treated the class operator $\Ca$ as
a ``matrix'' and not taken its complex conjugate to apply the inner
product $\circ$.}
\widetext
\begin{equation}\label{melt}
\langle\varphi_i|C_\a|\psi_j\rangle = \int_{\sigma ''} d^D\!\Sigma
''{}^\mu
\int_{\sigma '} d^D\!\Sigma '{}^\nu 
\varphi_i^*(x'')i\nabboth{}''_\mu
\Ca
i\nabboth{}'_\nu\psi_j(x').
\end{equation}
\narrowtext

     Integrating over all paths gives the unrestricted propagator
\begin{equation}
\Cu
=2mi\Delta_F(x''\!\!-\!x'),
\end{equation}
where
\begin{equation}
\Delta_F(x''\!\!-\!x')=\int{d^{D+1}p\over(2\pi)^{D+1}}
{e^{ip\cdot(x''-x')}\over -(p^2+m^2)+i\varepsilon}
\end{equation}
is the Feynman propagator,
which propagates positive energy solutions forward in time and
annihilates negative energy solutions:
\begin{mathletters}\label{feynpos}
\begin{eqnarray}
\Cu\circ e^{-i\op t'}e^{i\p\cdot \x'} 
&=& 2m e^{-i\op t''}e^{i\p\cdot \x''}\\
\Cu\circ e^{i\op t'}e^{i\p\cdot \x'}&=&0
\end{eqnarray}
\end{mathletters}
assuming $t''>t'$ (where $\op=\sqrt{{\bf p}^2+m^2}$).  The restriction
of the multiplier $N$ to positive values has given the advertized
correspondence to field theory, as our propagator is the familiar
Feynman propagator.  This has also led to the bias towards positive
energy solutions (\ref{feynpos}).

\subsection{Spacetime alternatives}

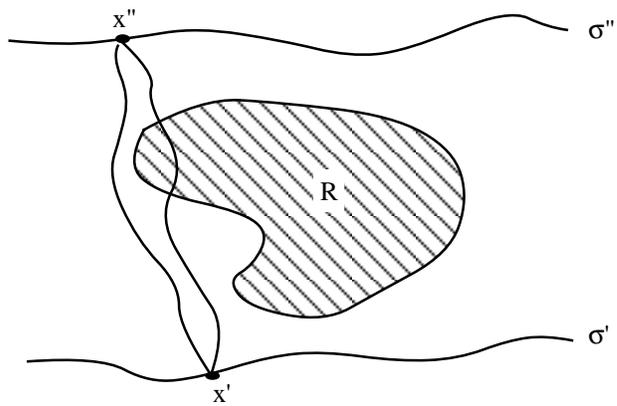
\begin{figure}
\caption{An example of a spacetime coarse graining.  The path on the left
never enters the spacetime region $R$ and is thus in the class
$c_\rb$.  The path on the right spends part of its trajectory in $R$
and is thus in the class $c_r$.  ($D-1$ of the $D$ space dimensions
have been suppressed.)}
\label{stalt}
\end{figure}

As an example of a simple spacetime coarse graining, we define a
spacetime region $R$, and a set of two exclusive and exhaustive
alternatives as follows: $c_r$ is the class of paths which at some
point enter $R$, and $c_\rb$ is the class of paths which never enter
it.  (See Fig.~\ref{stalt}.)  If we define
\begin{eqnarray}
&&\CrbN\nonumber\\
&&=\int\limits_{x'\rb x''}\D x\,\D p
\exp\left[i\int\nolimits_0^N d\tau \left(p\cdot{dx\over
d\tau}-{p^2+m^2\over 2m}\right)\right],\nonumber\\
\label{CrbN}
\end{eqnarray}
so that
\begin{equation}
\Crb=\int\nolimits_0^\infty dN \CrbN,
\end{equation}
comparing (\ref{CrbN}) to the path integral expression for a 
 nonrelativistic propagator, we can show (see\cite{leshouches}
for more details) that $\CrbN$ obeys a five-dimensional
Schr\"{o}dinger-like equation
\begin{mathletters}\label{PDEIC}
\begin{equation}\label{PDE}
\left(-i{\partial\over\partial N}+{-\nabla_{x''}^2+m^2\over 2m}
-iE_R(x'')\right)\CrbN=0
\end{equation}
with initial condition
\begin{equation}
\langle x''0||C_\rb||x'0\rangle
=\delta^{D+1}(x''\!\!-\!x')e^{-E_R(x')},
\end{equation}
\end{mathletters}
where we explicitly allow for the possibility
that the region $R$ intersects the initial slice $\sigma '$ or
the final slice 
$\sigma ''.$
Here 
\begin{equation}
E_R(x)=\cases{0,&$x\notin R$\cr
\infty,&$x\in R$\cr}
\end{equation}
is the excluding potential for the region $R$.  Note that
\begin{equation}
e^{-E_R(x)}=\cases{1,&$x\notin R$\cr 0,&$x\in R$\cr}.
\end{equation}

(\ref{PDEIC}) is equivalent to the homogeneous PDE
\begin{mathletters}\label{homog}
\begin{equation}\label{homogeqn}
\left(-i{\partial\over\partial N}+{-\nabla_{x''}^2+m^2\over 2m}
\right)\CrbN=0,\quad x''\notin R 
\end{equation}
with boundary condition
\begin{equation}\label{bc}
\CrbN=0,\quad
x''\in\partial R
\end{equation} and initial condition 
\begin{equation}
\langle
x''0||C_\rb||x'0\rangle=\delta^{D+1}(x''\!\!-\!x')e^{-E_R(x')}.
\end{equation}
\end{mathletters}

\section{Solution by method of images}\label{images}
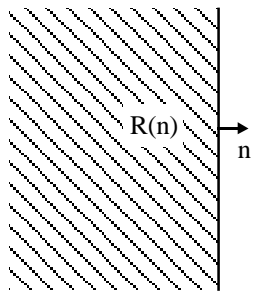
\begin{figure}
\caption{The region $R(n)$ defined by the unit vector $n$.  ($D-1$ of
the $D$ space dimensions have been suppressed.)}
\label{region}
\end{figure}

     For a sufficiently simple region, we can construct the class
operator $C_\rb$ by the method of images. Let $n$ be a constant
spacelike unit vector $(n\cdot n=1)$, and $x_n=n\cdot x$ be
the component of $x$ along $n$.  Then define $R(n)$ by $x_n\le 0$
(Fig.~\ref{region}), so that $e^{-E_{R(n)}(x)}=\Theta(x_n)$ (where
$\Theta$ is the Heavyside step function).  If we define\footnote{To
avoid confusion, keep in mind that $x_n$ is just a number, while $x_c$
is a $(D+1)$-vector.} the reflection of $x$ through the plane $x_n=0$
by $x_c=x-2 x_n n$, $\langle x''N||C_u||x'0\rangle-\langle
x''N||C_u||x_c'0\rangle$ satisfies (\ref{homogeqn}) (by the
principle of superposition) and (\ref{bc}), and has initial value
\begin{eqnarray}
\lefteqn{\langle x''0||C_u||x'0\rangle-\langle
x''0||C_u||x_c'0\rangle}\qquad\qquad\qquad\qquad\nonumber\\
&&=\delta^{D+1}(x''\!\!-\!x')-\delta^{D+1}(x''\!\!-\!x_c'),
\end{eqnarray}
which is equal to $\delta^{D+1}(x''\!\!-\!x')$ for
$x',x''\notin {R(n)}$.  Thus
\begin{eqnarray}
&&\CrnbN\nonumber\\
&&=\Theta(x'_n)\Theta(x''_n)
\left(\langle x''N||C_u||x'0\rangle-\langle x''N||C_u||x_c'0\rangle\right)
\end{eqnarray}
solves (\ref{homog}), and yields the class operator
\begin{eqnarray}
&&\Crnb\nonumber\\
&&=2mi\Theta(x'_n)\Theta(x''_n)
\left[\Delta_F(x''\!\!-\!x')-\Delta_F(x''\!\!-\!x'_c)\right].
\label{classmoi}
\end{eqnarray}

\section{Dependence on initial and final time slices}\label{surf}
     Since our construction (\ref{melt}) of the matrix elements
$\left\{\langle\varphi_i|C_\a|\psi_j\rangle\right\}$ from the class
operator $\Ca$ makes explicit reference to a choice of
 nonintersecting spacelike surfaces $\sigma '$ and $\sigma ''$, those
matrix elements and hence the decoherence functional could, in
principle, depend on the choice of surfaces, and we would like to
determine what, if any, that dependence is.  Observe that for a given
surface $\sigma$ with normal vector $u$, the Klein-Gordon inner
product (\ref{inner}) on that surface depends only
on the values on $\sigma$ of the wave function $\psi$ and its first
normal derivative $u^\mu\nabla_\mu\psi$.  Thus the construction of the
decoherence functional (\ref{dec}) depends only on the values on
$\sigma''$ of $\varphi_i(x'')$ and $u''^\mu\nabla''_\mu\varphi_i(x'')$
and the values on $\sigma'$ of $\psi_j(x')$ and
$u'^\nu\nabla'_\nu\psi_j(x')$.  To discuss the behavior of the
decoherence functional under changes of $\sigma'$ or $\sigma''$, we
need to define how the wave functions $\varphi$ and $\psi$ vary off of
those surfaces, and we do so by requiring them to satisfy the
Klein-Gordon equation.

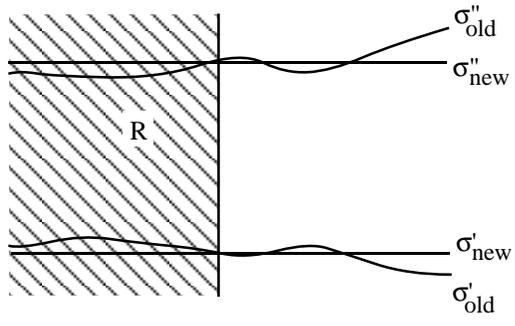
\begin{figure}
\caption{Varying the surfaces $\sigma'$ and $\sigma''$ on which the 
inner product (\protect\ref{inner}) is imposed does not change the
decoherence functional, as long as their intersections with $\partial
R$ are unchanged.  ($D-1$ of the $D$ space dimensions have been
suppressed.)}
\label{vary}
\end{figure}

     Now we can consider how
$\langle\varphi_i|C_\rb||x'\rangle=\varphi_i(x'')\circ\Crb$ varies
under changes of $\sigma ''$.  As a consequence of
(\ref{PDE}) the class operator $\Crb$ will satisfy the following (for
any region $R$):
\begin{mathletters}\label{kgoff}
\begin{eqnarray}\label{satisfyout}
\left({-\nabla_{x''}^2+m^2\over 2m}\right)\Crb=0,&\quad& x'\ne x''\notin
R\\
\label{vanishin}
\Crb=0,&\quad& x''\in R.
\end{eqnarray}
\end{mathletters}
We assume here, as throughout this work, that the surfaces $\sigma'$
and $\sigma''$ do not intersect one another, so that $x'\ne x''$ holds
as far as we're concerned.  Thus $\Crb$ satisfies the Klein-Gordon
equation on $x''$ everywhere except on the boundary $\partial R$.
Since the final wave functions $\{\varphi_i\}$ are taken to be
solutions to the Klein-Gordon equation, the usual demonstration of
invariance of the Klein-Gordon inner product tells us that we can
deform the surface $\sigma ''$ without changing
$\langle\varphi_i|C_\rb||x'\rangle$ so long as its intersection
$\sigma ''\cap\partial R$ with the boundary of R stays fixed.
Examining the behavior of the sum-over-histories construction
(\ref{class}) under the substitutions $\upsilon=N\!-\!\tau$,
$y(\upsilon)=x(N\!-\!\upsilon)$ and $k=-p$, we see that the class
operator is symmetric under the interchange of ends of the path
($\Ca=\langle x'||C_\a||x''\rangle$) so long as the class $c_\a$ does
not distinguish one end of the path from the other.  The class $c_\rb$
is such a class.\footnote{An example of a class which {\em does}
distinguish one end of the class from the other is one which refers to
the first time in its trajectory that a particle crosses a surface or
enters a region.} Thus $\Crb$ must satisfy the analogous properties to
(\ref{kgoff}) with respect to the other argument $x'$.  Thus changes
of $\sigma '$ which leave $\sigma '\cap\partial R$ unchanged will not
change $\langle x''||C_\rb|\psi_j\rangle=\Crb\circ\psi_j(x')$ either.
Since $\Cr +\Crb=\Cu=2mi\Delta_F(x''\!\!-\!x')$ by (\ref{classone}),
and the Feynman propagator satisfies the Klein-Gordon equation on its
(nonvanishing) argument, $\Cr$ will satisfy the equation whenever
$\Crb$ does, and all elements of the decoherence functional will be
unchanged under any change of $\sigma '$ and $\sigma ''$ which leaves
their intersection with $\partial R$ unchanged.  (Fig.~\ref{vary})

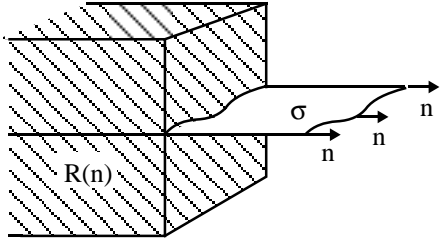
\begin{figure}
\caption{Generating the surface $\sigma$ from its intersection with
$\partial R(n)$ by projecting along $n$.  ($D-2$ of the $D$ space
dimensions have been suppressed.)  If $D=1$, $\sigma\cap\partial R(n)$
is a point and $\sigma$ generated in this fashion will always be flat.
With two or more space dimensions, $\sigma$ will only be flat if
$\sigma\cap\partial R(n)$ is; if $\sigma\cap\partial R(n)$ is
``wavy'', $\sigma$ will be translationally invariant along $n$,
resembling a sheet of corrugated metal.}
\label{project}
\end{figure}

\label{surfcons}
     This argument has previously been used\cite{leshouches} to
show that the decoherence functional is independent of the choice
of  nonintersecting surfaces so long as $\sigma '$ lies
completely to the past and $\sigma ''$ completely to the future
of $R$.  The nature of the region $R(n)$ defined in Sec.~\ref{images}
prevents us from choosing initial and final
spacelike surfaces which do not intersect $R(n)$.  What we can do
without changing the decoherence functional is generate the $D$-surface
$\sigma$ from
the ($D-1$)-surface $\sigma\cap\partial R(n)$ via curves everywhere
tangent to $n$.  (Fig.~\ref{project}) Then $n$ will lie in the surface
at all points, and
$n^\mu d^D\!\Sigma_\mu=0.$  \footnote{Note that it $1+1$
dimensions, this allows us to choose our surface to be a surface
of constant time in the reference frame where $n^0=0$.} This will
later prove crucial.

\section{Our chosen set of alternatives}\label{seclrb}
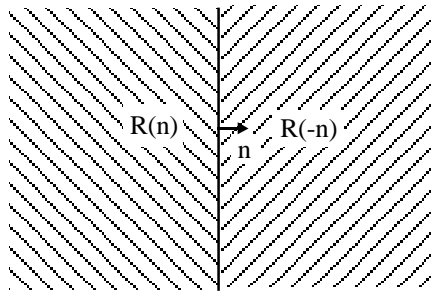
\begin{figure}
\caption{The regions $R(n)$ (``left'') and $R(-n)$ (``right'') defined
by the unit vector $n$, along with their common boundary, the ``wall''
$x_n=0$.  ($D-1$ of the $D$ space dimensions have been suppressed.)}
\label{twosides}
\end{figure}

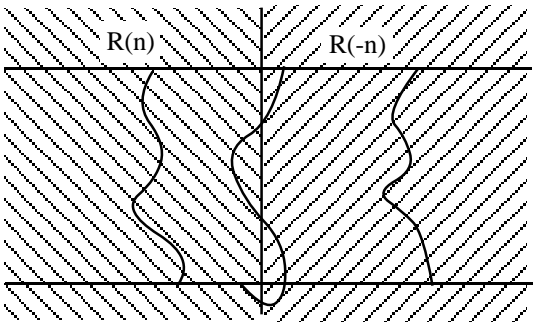
\begin{figure}
\caption{The coarse graining described in Sec.~\protect\ref{seclrb}.
The three paths shown are representatives of, from left to right: the
class $c_l$ of paths which lie completely to the left of the wall; the
class $c_b$ of paths which spend some time on each side of the wall;
and the class $c_r$ of paths which lie completely to the right of the
wall.  ($D-1$ of the $D$ space dimensions have been suppressed.)
Compare Fig.~3 of \protect\cite{nonrel}.}
\label{lrb}
\end{figure}

     We can take advantage of the fact that for a given normal vector
$n$, the regions $R(n)$ ($n\cdot x\le 0$) and $R(-n)$ ($-n\cdot x\le
0$) are on opposite sides of the same boundary $x_n=0$.
(Fig.~\ref{twosides}) Loosely calling $R(n)$ the ``left'' side and
$R(-n)$ the ``right'' side of the ``wall'' $x_n=0$, we can define a
set of alternatives by the answers to the two questions ``does the
particle ever enter $R(n)$ ($x_n\le 0$)?''\ and ``does the particle
ever enter $R(-n)$ ($x_n\ge 0$)?''  The class $c_\rnb\cap c_{\overline
r(-n)}$, corresponding to both answers being ``no'', is empty.  The
three nontrivial alternatives are: $c_l=c_{r(n)}\cap c_{\overline
r(-n)}=c_{\overline r(-n)}$, in which the particle is on the left side
of the wall throughout its entire trajectory; $c_r=c_\rnb\cap
c_{r(-n)}=c_\rnb$, in which the particle is always on the right side;
and $c_b=c_{r(n)}\cap c_{r(-n)}$, in which the particle spends some
time on each side of the wall, and crosses it in between.  This set of
three alternatives, illustrated in Fig.~\ref{lrb}, is exhaustive and
mutually exclusive, and is thus a suitable coarse graining.  The class
operators for $c_l$ and $c_r$ were calculated in Sec.~\ref{images},
and are given by
\begin{mathletters}\label{Clrb}
\begin{eqnarray}
&&\Cl=\langle x''||C_{\overline r(-n)}||x'\rangle\nonumber\\
&&=2mi\Theta(-x'_n)\Theta(-x''_n)
\left[\Delta_F(x''\!\!-\!x')-\Delta_F(x''\!\!-\!x'_c)\right]\nonumber\\
\end{eqnarray}
\begin{eqnarray}
&&\Cr=\langle x''||C_{\overline r(n)}||x'\rangle\nonumber\\
&&=2mi\Theta(x'_n)\Theta(x''_n)
\left[\Delta_F(x''\!\!-\!x')-\Delta_F(x''\!\!-\!x'_c)\right],
\end{eqnarray}
where we have used the fact that $x_{-n}=-n\cdot x=-x_n$ [and also that
$x_c$ is defined the same way with respect to $n$ and $-n$: $x_c=x-2 n
x_n=x+2 n x_{-n}=x-2(-n)x_{-n}$].  The class
operator for $c_b$ can be calculated from the superposition law
(\ref{classone}):
\begin{eqnarray}
&&\Cb=\Cu-\Cl-\Cr\nonumber\\
&&=2mi\{[\Theta(x'_n)\Theta(-x''_n)
+\Theta(-x'_n)\Theta(x''_n)]
\Delta_F(x''\!\!-\!x')\nonumber\\
&&\mbox{}-[\Theta(x'_n)\Theta(x''_n)+\Theta(-x'_n)\Theta(-x''_n)]
\Delta_F(x''\!\!-\!x'_c)\}.
\end{eqnarray}
\end{mathletters}

\section{Properties for certain initial and final conditions}
\subsection{Pure initial state}\label{pure}
     If we specialize to a pure initial state $\psi(x')$, it
becomes useful to define the branch wave function
\begin{equation}\label{branch}
\psi_\a (x'')={1\over 2m}\langle x''||C_\a|\psi\rangle={1\over
2m}\Ca\circ\psi(x'),
\end{equation}
so that the decoherence functional (\ref{dec}) has elements
\begin{equation}\label{puredec}
D(\a,\ap)={\psi_{\ap}\circ\rho
''\circ\psi_{\a}\over\psi^+\circ\rho ''\circ\psi^+}.
\end{equation}
Here $\psi^+$ is the positive energy part of $\psi$ [see
(\ref{feynpos})]:
\begin{equation}
\psi^+(x'')=i\Delta_F(x''\!\!-\!x')\circ\psi(x')=
{1\over 2m}\langle x''||C_u|\psi\rangle,
\end{equation}
and is the branch wave function corresponding to the class $c_u$ of all
paths.  The superposition property for class operators
(\ref{classsuper}) and the definition of the branch wave function
(\ref{branch}) imply an analogous superposition law for branch
wave functions:
\begin{mathletters}
\begin{eqnarray}\label{branchsuper}
&&\psi_\beta(x'')=\sum_{\a\in\beta}\psi_\a(x'')\\
&&\sum_\a\psi_\a(x'')=\psi^+(x'').
\end{eqnarray}
\end{mathletters}

     We postpone for the moment discussion of the final condition
$\rho ''$.

     The branch wave functions for the classes $c_l$, $c_r$ and $c_b$
can be given in terms of the branch wave functions $\psi_{\rb(\pm n)}$
by
\begin{mathletters}
\begin{eqnarray}
\psi_l(x'')&=&\psi_{\rb(-n)}(x'')\\
\psi_r(x'')&=&\psi_{\rb(n)}(x'')\\
\psi_b(x'')&=&\psi^+(x'')-\psi_l(x'')-\psi_r(x'').
\end{eqnarray}
\end{mathletters}

     Using (\ref{classmoi}), we write $\psi_\rpmnb(x'')$ as
\widetext
\begin{equation}
\psi_\rpmnb(x'')=\Theta(\pm x''_n)\int_{\sigma'}d^D\!\Sigma'{}^\nu
\Theta(\pm x'_n)
\left[i\Delta_F(x''\!\!-\!x')-i\Delta_F(x''\!\!-\!x'_c)\right]
i\nabboth{}'_\nu\psi(x').
\end{equation}
As described in Sec.~\ref{surfcons}, we can, without loss of
generality, choose $\sigma '$ to satisfy $n_\nu d^D\!\Sigma'{}^\nu=0$,
which allows us to move the $\Theta(\pm x'_n)$ to the other side of
the $\nabboth{}'_\nu$ [since $\nabla_\nu\Theta(\pm x'_n)=
\pm n_\nu\delta(x'_n)$, which is orthogonal to $d^D\!\Sigma'{}^\nu$] and
get
\begin{equation}\label{branch-1}
\psi_\rpmnb(x'')=\Theta(\pm x''_n)\int_{\sigma'}d^D\!\Sigma'{}^\nu
\left[i\Delta_F(x''\!\!-\!x')-i\Delta_F(x''\!\!-\!x'_c)\right]
i\nabboth{}'_\nu\Theta(\pm x'_n)\psi(x').
\end{equation}
If we change the integration variable from $x'$ to $x'_c$ in the
second term of the integral
(which we can do because the construction of $\sigma'$ ensures
that $x'_c\in\sigma'$ if and only if $x'\in\sigma'$), we obtain
\begin{equation}\label{branchpsi}
\psi_\rpmnb(x'')=\Theta(\pm x''_n)\int_{\sigma'}
d^D\!\Sigma'{}^\nu i\Delta_F(x''\!\!-\!x')i\nabboth{}'_\nu
\left[\psi(x')\Theta(\pm x'_n) -
\psi(x'_c)\Theta(\mp x'_n)\right]
\end{equation}
\narrowtext
Without an additional restriction on $\psi(x')$, it is quite
difficult to proceed any further.

\subsubsection{Antisymmetric initial state}\label{antisym}
     If we choose our initial state to be an odd function of $x_n$
(which we write as $\chi$ to distinguish it from the generic initial
state $\psi$):
\begin{equation}\label{chi}
\chi(x_c)=-\chi(x),
\end{equation}
 we have
$\chi(x')\Theta(\pm x'_n) - \chi(x'_c)\Theta(\mp x'_n)=\chi(x')$,
and (\ref{branchpsi}) becomes
\begin{eqnarray}
\chi_\rpmnb(x'')&=&\Theta(\pm x''_n)\int_{\sigma'}
d^D\!\Sigma'{}^\nu i\Delta_F(x''\!\!-\!x')
i\nabboth{}'_\nu\chi(x')\nonumber\\
&=&\Theta(\pm x''_n)\chi^+(x'').
\end{eqnarray}

     Thus the branch wave functions for this initial state are
\begin{mathletters}
\begin{eqnarray}
\chi_l(x'')&=&\chi_{\overline r(-n)}(x'')=\Theta(-x''_n)\chi^+(x'')
\label{chil}\\
\chi_r(x'')&=&\chi_\rnb(x'')=\Theta(x''_n)\chi^+(x'')
\label{chir}\\
\chi_b(x'')&=&0.
\label{chib}
\end{eqnarray}
\end{mathletters}

     Note that we can construct a Klein-Gordon state satisfying the
antisymmetry property (\ref{chi}) throughout all spacetime by taking
any Klein-Gordon state $\zeta(x)$ which is not symmetric about $x_n=0$
and defining $\chi(x)={1\over 2}\left[\zeta(x)-\zeta(x_c)\right]$, and
note also that both the positive and negative energy parts of $\chi$
have the antisymmetry property as well.

\subsubsection{Initial state with restricted support}\label{support}
     Another technique for simplifying the branch wave functions,
used on the  nonrelativistic particle by Yamada and Takagi
\cite{Y&T} is to choose an initial state which vanishes either in
or out of the region $R$.  Since we attach the initial state with the
Klein-Gordon inner product, we need to go a step further, and require
that both the initial state $\psi(x')$ and its normal derivative
$u'{}^\nu\nabla'_\nu\psi(x')$ vanish on the appropriate part of the
initial surface.  For brevity's sake, we define the ``support'' of a
wave function to be anywhere where the wave function or its normal
derivative is nonvanishing.  Thus we want to construct a wave function
whose support on the initial surface $\sigma'$ is confined to (say)
the left side of the wall ($x_n< 0$).  It is always possible to
construct a solution to the Klein-Gordon equation $\psi(x)$ which has
an arbitrary value $f(x')$ and normal derivative $g(x')$ on a surface
$\sigma'$, but it will in general be necessary to construct it out of
both positive and negative energy components.\footnote{I am indebted
to R.~S.~Tate for pointing this out to me.}

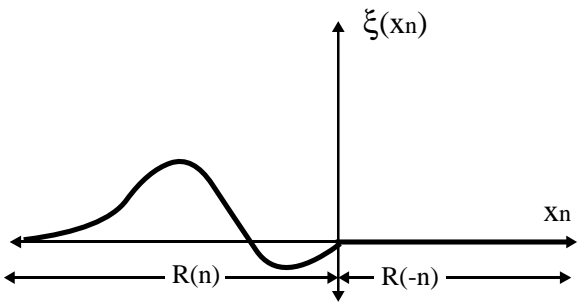
\begin{figure}
\caption{Schematic plot of a wave function $\xi$ whose support on
$\sigma'$ is confined to $x_n< 0$.  This is a plot of $\xi$ as a
function of $x_n$ for fixed $\text{\bf x}_\perp$ on the surface
$\sigma'$.  Note that $u'^\nu\nabla'_\nu\xi(x')$ must also vanish on
the ``right'' half of the surface $\sigma$ for $\xi$ to have
semi-infinite support as defined in Sec.~\protect\ref{support}.  [See
(\protect\ref{eqnxi}).]}
\label{xi}
\end{figure}

     If we construct an initial state (which we call $\xi$) whose
support on the surface $\sigma'$ is confined to the left side of the
wall:
\begin{equation}\label{eqnxi}
\xi(x')=0=u'\cdot\nabla'\xi(x') \hbox{ when } x'\in\sigma \hbox{
and }x'_n\ge 0
\end{equation}
(see Fig.~\ref{xi}), then
$\Theta(x'_n)\xi(x')$ and its normal derivative vanish
and (\ref{branch-1}) gives
\begin{mathletters}
\begin{equation}\label{xir}
\xi_r(x'')=\xi_\rnb(x'')=0.
\end{equation}

     Turning the tables and considering the effect the semi-infinite
support property (\ref{eqnxi}) has on $\xi_l=\xi_{\overline r(-n)}$,
we see that $\Theta(-x'_n)\xi(x')$ has the same value and normal
derivative on $\sigma'$ as $\xi$ itself, and we will be able to drop
the $\Theta(-x'_n)$ from (\ref{branch-1}), and obtain
\begin{equation}\label{xil}
\xi_l(x'')=\xi_{\overline
r(-n)}(x'')=\Theta(-x''_n)\left[\xi^+(x'')-\xi^+(x''_c)\right].
\end{equation}
[We have used the easily proved result that
$\Delta_F(x''\!\!-\!x'_c)=\Delta_F(x''_c\!\!-\!x')$.]

     $\xi_b$ can again be found by superposition, and is given by:
\begin{equation}\label{xib}
\xi_b(x'')=\Theta(x''_n)\xi^+(x'')+\Theta(-x''_n)\xi^+(x''_c).
\end{equation}
\end{mathletters}

\subsection{Future indifference}\label{future}
     In order to evaluate the decoherence functional (\ref{puredec})
we need to consider the final condition $\rho''$.  In analogy with our
observations that the universe has a preferred time direction, we
would like to abandon the time-symmetric construction of (\ref{dec})
and choose a condition of future indifference, i.e., a completely
unspecified final condition.  In most time-symmetric formulations of
quantum mechanics, this condition is implemented by replacing the
final density matrix with the identity operator, so that
$\psi_{\ap}\circ\rho ''\circ\psi_\a\rightarrow\psi_\ap\circ\psi_\a$,
but this cannot be the prescription here, since it is not manifestly
positive when $\a=\ap$, as our initial construction was.

To see why this fails, construct completely unspecified density
matrices for the positive and negative energy sectors of the theory:
\begin{equation}\label{rhom}
\rho_\pm(x_2,x_1)=\int {d^D\!p\over (2\pi)^D 2\op}e^{\mp i\op
(t_2-t_1)}e^{i\p\cdot (\x_2-\x_1)}.
\end{equation}
They have the following property under the Klein-Gordon inner
product:
\begin{equation}\label{pm}
\rho_\pm(x_2,x_1)\circ\psi(x_1)=\pm\psi^\pm(x_2),
\end{equation}
where $\psi(x)$ is any solution to the Klein-Gordon equation, and
$\psi^+(x)$ and $\psi^-(x)$ are its positive and negative energy
components, respectively [$\psi(x)=\psi^+(x)+\psi^-(x)$].  The
``identity operator'' with respect to this inner product is thus
$\rho_+-\rho_-$.  It is unsuitable for a final condition $\rho''$,
since some of the weights $\{p''_i\}$ it implies are negative,
in violation of the rules set out in Sec.~\ref{setup}.  Instead, we
take our condition of future indifference to be
\begin{equation}\label{rhofi}
\rho_{fi}=\rho_+ +\rho_-,
\end{equation}
so that\footnote{Technically speaking, we should not talk about the
positive and negative energy components of the branch wave functions
$\{\psi_\a\}$, since we showed in Sec.~\ref{surf} that the class
operators (and hence the branch wave functions) are guaranteed to
satisfy the Klein Gordon equation only when $x''\notin\partial R$, and
the branch wave functions are thus not in the space of solutions to
the Klein-Gordon equation.  However, a more careful analysis (see the
Appendix) shows that, defining $\psi^\pm$ by (\ref{pm}),
$\varphi\circ\psi =\varphi^+\circ\psi^+ + \varphi^-\circ\psi^-
=(\varphi^++\varphi^-) \circ (\psi^++\psi^-)$ (where all inner
products are taken on the same surface), even if $\varphi$ and $\psi$
are not solutions to the Klein-Gordon equation.  The division into
positive and negative energy parts is thus well-defined for our
purposes.}
\begin{equation}
\psi_{\ap}\circ\rho_{fi}\circ\psi_\a=
\psi^+_{\ap}\circ\psi^+_\a-\psi^-_{\ap}\circ\psi^-_\a.
\end{equation}
This is equivalent to the result we would have gotten if we had used
the positive definite inner product for Klein-Gordon wave functions,
and then chosen the identity as our final density matrix.  This inner
product is nonlocal in the spacetime co\"{o}rdinate $x$, so, for
example, wave functions which do not overlap can still have a
nonvanishing inner product.

     Note that we can replace the normalization factor
$\psi^+\circ\rho''\circ\psi^+$ in (\ref{puredec}) with
$\psi^+\circ\psi^+$ if we use the final condition (\ref{rhofi}).  It
will therefore prove useful to normalize our initial wave function so
that
\begin{equation}\label{norm}
\psi^+\circ\psi^+=1.
\end{equation}
The decoherence functional is then
\begin{equation}\label{normal}
D(\a,\ap)=\psi_{\ap}\circ\rho_{fi}\circ\psi_{\a}.
\end{equation}

\section{Results}

\subsection{Results for antisymmetric initial state}\label{antisymres}
     Using the antisymmetric initial state $\chi$ from
Sec.~\ref{antisym}, the branch wave functions for the three classes
are
\begin{mathletters}
\begin{eqnarray}
\chi_l(x'')&=&\Theta(-x''_n)\chi^+(x'')
\eqnum{\protect\ref{chil}}\\
\chi_r(x'')&=&\Theta(x''_n)\chi^+(x'')
\eqnum{\protect\ref{chir}}\\
\chi_b(x'')&=&0.
\eqnum{\protect\ref{chib}}
\end{eqnarray}
\end{mathletters}
The elements of the decoherence functional (\ref{normal}) are
calculated in the Appendix, and found (when the final surface
$\sigma''$ is taken to be one of constant time $t''$) to be
\begin{mathletters}\label{result}
\begin{equation}\label{Dchi}
\left(
\begin{array}{lll}
D(l,l)={1\over 2}+\Delta D & D(l,r)=-\Delta D & D(l,b)=0 \\
D(r,l)=-\Delta D & D(r,r)={1\over 2}+\Delta D & D(r,b)=0 \\
D(b,l)=0 & D(b,r)=0 & D(b,b)=0 \\
\end{array}\right)
\end{equation}
where\footnote{We use here several pieces of notation defined in the
Appendix, namely ${\bf v_\perp= \bf v}-v_n{\bf n}$ and
$\omega_\perp=\sqrt{{\bf k}_\perp^2+m^2}$, and also that
$\tilde\chi^+$ is the Fourier transform (\ref{Fourier}) of the
positive energy part of $\chi$.  We are also working in a reference
frame where ${\bf n}$ has no time component.}
\widetext
\begin{equation}\label{dD}
\Delta D=
2\int{dk_{1n}dk_{2n}d^{D-1}k_\perp\over (2\pi)^2}
{\omega_1+\omega_2\over 2\sqrt{\omega_1\omega_2}}
\tilde\chi^+({\bf k}_2)^*\tilde\chi^+({\bf
k}_1){e^{-i(\omega_1-\omega_2)t''}
\over k_{1n}-k_{2n}}
\ln\left({\omega_1-k_{1n}\over\omega_2-k_{2n}}\right).
\end{equation}
\end{mathletters}
\narrowtext
Aside from $D(l,r)=D(r,l)=-\Delta D=\chi_l\circ\rho_{fi}\circ\chi_r$,
all of the off-diagonal elements vanish (this is true for any final
condition, in fact).  $D(l,r)=D(r,l)$ generally does not vanish,
despite the lack of overlap of the branch wave functions, because of
the nonlocality of the positive definite inner product induced by the
final condition in section \ref{future}.  Note that whenever the
alternatives do decohere ($\Delta D\approx 0$), the probabilities are
given by $p(l)\approx 1/2 \approx p(r)$, $p(b)=0$.  [Symmetry
arguments make it clear that we must have $p(l)=p(r)$.]  Note also
that while the decoherence functional depends on the time $t''$ of the
final surface, it is completely independent of the initial surface
$\sigma'$.

     To determine
whether or not we have decoherence, we need to consider further
properties of the initial condition $\chi$ (or equivalently its
Fourier transform $\tilde\chi$).

     Let $\tilde\chi$ be given by a Gaussian wavepacket peaked at
${\bf k}_0$, ${\bf x}_0$ and $t_0$, minus its reflection through
$k_n=0$.  That is to say
\widetext
\begin{mathletters}
\begin{eqnarray}
\tilde\chi({\bf k})
&=&Ce^{i\ok t_0}\left(e^{-i\k\cdot\xo}e^{-(\k-\ko)^2/4(\delta k)^2}
-e^{-i\kc\cdot\xo}e^{-(\kc-\ko)^2/4(\delta k)^2}\right)
\nonumber\\
&=&Ce^{-i\kp\cdot\xo}e^{i\ok t_0}e^{-(\kp-\kop)^2/4(\delta k)^2}
\sum_{\xi=\pm 1}\xi e^{-i\xi k_n x_{0n}}
e^{-(k_n-\xi k_{0n})^2/4(\delta k)^2},
\label{Gaussian}
\end{eqnarray}
where the normalization constant is given by
\begin{equation}
|C|^2={1\over 2 (\delta k\sqrt{2\pi})^D
\left[1-e^{-k_{0n}^2/2(\delta k)^2}e^{-x_{0n}^2/2(\delta x)^2}\right]}
\end{equation}
\end{mathletters}
with $\delta x\delta k =1/2$.  We then have
\begin{eqnarray}
\Delta D&=&2|C|^2\int{dk_{1n}dk_{2n}d^{D-1}k_\perp\over (2\pi)^2}
{\omega_1+\omega_2\over 2\sqrt{\omega_1\omega_2}}
e^{-(\kp-\kop)^2/2(\delta k)^2}e^{-i(\omega_1-\omega_2)(t''-t_0)}
\ln\left({\omega_1-k_{1n}\over\omega_2-k_{2n}}\right)
\nonumber\\
&&\times
\sum_{\xi_1=\pm 1}\sum_{\xi_2=\pm 1}\xi_1\xi_2 
e^{-(k_{1n}-\xi_1 k_{0n})^2/4(\delta k)^2}
e^{-(k_{2n}-\xi_2 k_{0n})^2/4(\delta k)^2}
{e^{-i\xi_1 k_{1n} x_{0n}}e^{i\xi_2 k_{2n} x_{0n}}
\over k_{1n}-k_{2n}}
\nonumber\\
&=&2|C|^2\int{dk_{1n}dk_{2n}d^{D-1}k_\perp\over (2\pi)^2}
{\omega_1+\omega_2\over 2\sqrt{\omega_1\omega_2}}
e^{-(\kp-\kop)^2/2(\delta k)^2}
e^{-(k_{1n}- k_{0n})^2/4(\delta k)^2}
e^{-(k_{2n}- k_{0n})^2/4(\delta k)^2}
\nonumber\\
&&\times
e^{-i(\omega_1-\omega_2)(t''-t_0)}
e^{-i (k_{1n}-k_{2n}) x_{0n}}
\sum_{\xi=\pm 1}{2\xi\over k_{1n}-\xi k_{2n}}
\ln\left({\omega_1-k_{1n}\over\omega_2-\xi k_{2n}}\right),
\end{eqnarray}
where the final form has been arrived at by changing the variables in
the integrals $k_{1n}\rightarrow\xi_1 k_{1n}$, $k_{2n}\rightarrow\xi_2
k_{2n}$ and then making the substitution $\xi=\xi_1\xi_2$.

     In the limit that $\delta k\rightarrow 0$, we can replace
$k_{1n}$ and $k_{2n}$ with $k_{0n}$ and ${\bf k}_\perp$ with 
${\bf k}_{0\perp}$ everywhere except in the Gaussian factors and
perform the integrals.  We can do this because
\begin{equation}
\lim_{k_{1n}\rightarrow k_0} {1\over k_{1n}- k_{0n}}
\ln\left({\omega_1-k_{1n}\over\omega_0- k_{0n}}\right)
=-{1\over \omega_0}
\end{equation}
is finite, and we obtain
\begin{eqnarray}
\Delta D&=&{2|C|^2\over (2\pi)^2} (\delta k\sqrt{4\pi})^2
 (\delta k\sqrt{2\pi})^{D-1}\,
2\left[-{1\over \omega_0} + {1\over k_{0n}}
\ln\left({\omega_0-k_{0n}\over\omega_{0\perp}}\right)\right]
+{\cal O}([\delta k]^2)
\nonumber\\
&=&-4{\delta k\over (2\pi)^{3/2}}\left[{1\over \omega_0} -
{1\over k_{0n}}\ln\left({\omega_0-k_{0n}\over\omega_{0\perp}}\right)\right]
+{\cal O}([\delta k]^2).
\end{eqnarray}
\narrowtext
Thus we have approximate decoherence to lowest order in $\delta k$.
Note that the first order correction to the decoherence functional is
independent of the time $t''$ of the final surface.

     For a generic antisymmetric initial condition $\chi$, (\ref{dD})
has no reason to be small, so the current set of alternatives will
probably not decohere.  However, consider a coarser graining in which
$c_l$ and $c_r$ are combined into a single class $c_o$, consisting of
all paths which stay on one side or the other of the wall, and never
cross it.  We can use the superposition property (\ref{super}) to
construct the decoherence functional from the finer-grained one
(\ref{Dchi}).  $D(o,o)=D(l,l)+D(l,r)+D(r,l)+D(r,r)=1$, etc.

The elements of the decoherence functional are given by
\begin{equation}
\left(
\begin{array}{ll}
D(o,o)=1 & D(o,b)=0 \\
D(b,o)=0 & D(b,b)=0 \\
\end{array}\right)
\end{equation}
so we have exact decoherence, and probabilities of 1 for $c_o$ and 0
for $c_b$.  This corresponds to the definite prediction that for
a pure initial state antisymmetric about $x_n=0$, the
particle path will not cross that surface.  Since the
antisymmetry property holds throughout all spacetime, this result
is independent of the choice of initial and final surfaces.

     This last result can be seen from another point of view, allowing
a slight generalization.  Using the superposition property for branch
wave functions (\ref{branchsuper}), we can construct
\begin{equation}
\chi_o(x'')=\chi_l(x'')+\chi_r(x'')=\chi^+(x'').
\end{equation}
Recalling that
\begin{equation}
\chi_b(x'')=0\eqnum{\protect\ref{chib}},
\end{equation}
we see that all branch wave functions but one vanish.  Examination of
(\ref{puredec}) shows that whenever this is the case, the only
nonvanishing element of the decoherence functional will be the
diagonal one corresponding to the alternative with the nonvanishing
branch wave function, and we will have decoherence, and a definite
prediction of that alternative.  This will hold for any final
condition [except of course for pathological cases when the final
condition is inconsistent with the initial condition
($\psi\circ\rho''\circ\psi=0$), in which case the denominator of
(\ref{puredec}) vanishes, and the decoherence functional is
ill-defined].

\subsection{Results for initial state with restricted support}
     With the initial state $\xi$ from Sec.~\ref{support},
which vanishes, along with its normal derivative, on the surface
$\sigma'\cap R(-n)$, we find that the branch wave functions for
the three classes are
\begin{eqnarray}
\xi_l(x'')&=&\Theta(-x''_n)[\xi^+(x'')-\xi^+(x''_c)]
\eqnum{\protect\ref{xil}}\\
\xi_r(x'')&=&0
\eqnum{\protect\ref{xir}}\\
\xi_b(x'')&=&\Theta(x''_n)\xi^+(x'')+\Theta(-x''_n)\xi^+(x''_c).
\eqnum{\protect\ref{xib}}
\end{eqnarray}
Now the wave functions $\xi_l$ and $\xi_b$ overlap, so we do not
expect decoherence, even na\"{\i}vely, unless we coarse grain so
that only one of the branch wave functions is  nonvanishing.  This
amounts to recombining $c_l$ and $c_b$ into $c_{r(n)}$, so that
the decoherence functional is
\begin{equation}
\left(
\begin{array}{ll}
D\bbox(r(n),r(n)\bbox)=1 & D\bbox(r(n),\rnb\bbox)=0 \\
D\bbox(\rnb,r(n)\bbox)=0 & D\bbox(\rnb,\rnb\bbox)=0 \\
\end{array}\right)
\end{equation}
which decoheres, with probabilities of 1 for $c_{r(n)}$ and 0 for
$c_\rnb$.  Here we have a definite prediction that the particle
will at some point in its trajectory be found in $R(n)$.  This
result, however, depends very much on the choice of the initial
surface $\sigma'$.

\section{Discussion}

     For our simple coarse graining (see Fig.~\ref{lrb}), we were able
to calculate explicit expressions for the class operators $C_{r(\pm
n)}$ and $C_{\rb(\pm n)}$, and hence for $C_l$, $C_r$ and $C_b$.

     To calculate branch wave functions for a pure initial state, we
chose the state to satisfy special conditions.
\begin{itemize}
\item{If the wave function $\chi$ was antisymmetric under
reflection through $x_n=0$, the branch wave function $\chi_b$ vanished,
while the  nonvanishing branches $\chi_l$ and $\chi_r$ had no overlap.
This result held no matter what the initial surface $\sigma'$.}
\item{If the wave function $\xi$ and its first normal derivative
vanished on that part of the initial surface $\sigma'$ which was
outside of $R(n)$, the branch wave function $\xi_r$ vanished, but the
other two branches, $\xi_l$ and $\xi_b$, overlapped.  This held only
for one specific choice of $\sigma'$}
\end{itemize}

     We could not simply take the inner product of branch wave
functions to calculate the decoherence functional, since that would
have been tantamount to choosing a non-positive-definite final density
matrix.  Thus even for the initial state $\chi$, the alternatives
$c_l$ and $c_r$ did not automatically decohere just because the branch
wave functions did not overlap.  If we restricted the final surface to
be flat, we could calculate explicit expressions for the elements of
the decoherence functional.  For some choices of initial state, the
off-diagonal elements were small, but in general they could be
appreciable.  Whenever the alternatives did decohere, the probability
for each was $1/2$, which we would have predicted on symmetry grounds.

     If we coarser grained either example so that only one branch
wave function was  nonvanishing, we of course found decoherence and a
definite prediction (probability 1) of the other alternative,
viz.:
\begin{itemize}
\item{For the initial condition $\chi$, if the alternatives were chosen
to be $c_b$ and $c_o=c_l\cup c_r$, we found decoherence for any
nonpathological final condition, with probabilities of 0 and 1,
respectively.  This was a definite prediction that the particle did
not cross $x_n=0$, given an antisymmetric initial condition.}
\item{For the initial condition $\xi$, if the alternatives were chosen
to be $c_\rnb=c_r$ and $c_{r(n)}=c_l\cup c_b$, we found decoherence
for any nonpathological final condition, with probabilities of 0 and
1, respectively.  This was a definite prediction that the particle
spent part of its trajectory in $R(n)$, given an initial condition
which had no support outside of $R(n)$.  This is hardly surprising,
and it only holds if we attach the initial wave function on the
correct hypersurface.}
\end{itemize}

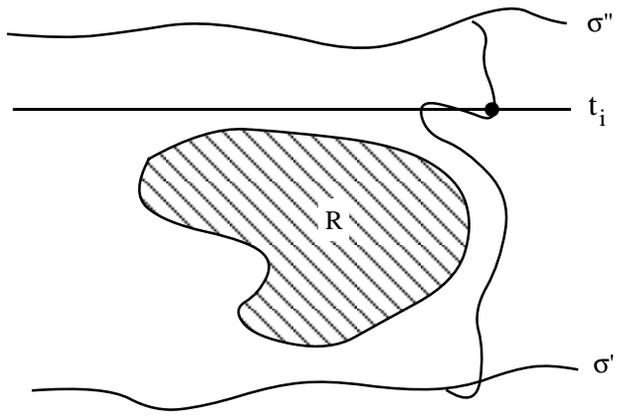
\begin{figure}
\caption{Dividing up a path which avoids a compact region $R$.
The path from $\sigma'$ to the last crossing of the intermediate
surface $t_i$ is in the class of paths from $\sigma'$ to $t_i$ which
avoid $R$.  The path from the last crossing of $t_i$ to $\sigma''$ is
in the class of paths from $t_i$ to $\sigma''$ which do not cross back
over $t_i$, and can be defined without reference to $R$.  ($D-1$ of
the $D$ space dimensions have been suppressed.)}
\label{lastcross}
\end{figure}

     Finally, let us observe that many of our complications were a
result of the fact the region which we considered intersected with our
initial and final surfaces.  If we had considered a region $R$ bounded
in time, we could have chosen our initial surface to lie to the past
and our final surface to the future of it.  As was discussed in
section \ref{surfcons}, this would make the decoherence functional
necessarily independent of the choice of surface.  It would also have
eliminated the complications in the choice of the final condition,
since the branch wave functions would have been positive energy
solutions to the Klein-Gordon equation.  The proof is straightforward:
construct an intermediate surface of constant time $t_i$ to future of
$R$ but the past of $\sigma''$.  (Section \ref{surfcons} always allows
us to deform the surface $\sigma''$ so that such a constant-time
surface will ``fit'' in.)  By a construction analogous to that of
Halliwell and Ortiz\cite{H&O}, the propagation from $\sigma'$ to
$\sigma''$ avoiding the region $R$ can be broken up (at the last
crossing of $t_i$) into propagation from $\sigma'$ to $t_i$ avoiding
$R$ followed by propagation from $t_i$ to $\sigma''$ which does not
cross back over $t_i$.  (See Fig.~\ref{lastcross}.)  The class
operator can thus be written
\begin{equation}
\Crb=\int d^D\! x_i \Delta_{1t_i}(x'',x_i)
\langle{\bf x}_i t_i||C_r|| x'\rangle
\end{equation}
where $\Delta_{1t_i}$ is the Newton-Wigner propagator:
\begin{equation}
\Delta_{1t_i}(x,x_i)=
\int {d^D\!p\over (2\pi)^D} e^{i\p\cdot(\x-\exi)} e^{-i\op(t-t_i)}.
\end{equation}
Since $\Delta_{1t_i}$ is constructed from positive-energy solutions of
the Klein-Gordon operator, the branch wave functions $\psi_r$ and
$\psi_\rb$ will each be positive-energy solutions themselves.  Thus
$\psi_\ap\circ\rho_-\circ\psi_\a=0$, so
$\psi_\ap\circ\rho_{fi}\circ\psi_\a=
\psi_\ap\circ\rho_+\circ\psi_\a= \psi_\ap\circ\psi_\a$, 
and we really do simply calculate the inner product of the branches.

     However, it was the simplicity of the region $R(n)$ which allowed
us to solve the PDE problem analytically in the first place.  Solution
of (\ref{homog}) for finite regions of spacetime cannot be
accomplished through straightforward method-of-images or
separation-of-variables methods.  In the nonrelativistic case, this
problem is circumvented for example in the case of a region which
extends from $t_1$ to $t_2$ by propagating from $t'$ to $t_1$ with the
free propagator, from $t_1$ to $t_2$ with the restricted propagator
calculated as though the region existed for all time, and then from
$t_2$ to $t''$ with the free propagator.  Since our paths are not
single-valued in time, we cannot ``turn off'' the restricting region
before and after we reach it, since we have to include in the sum
paths which double back into a previous regime.

\section{Conclusions}

     Using the generalized quantum mechanics formalism described by
Hartle for the quantum mechanics of the relativistic worldline, we
have examined one particularly simple coarse graining.  For a suitable
choice of initial conditions, albeit a more restrictive one than for
the nonrelativistic theory, we were able to assign approximate
probabilities to some sets of alternatives.

\acknowledgements

     The author wishes to thank K.~V.~Kucha\v{r} for asking to see a
summary of this project which evolved into the present work, N.~Yamada
and R.~S.~Tate for probing questions, D.~A.~Craig for comments on an
earlier draft, A.~Chamblin for helping to produce the figures, and
especially J.~B.~Hartle for advice, direction, and encouragement.
This material is based upon work supported under a National Science
Foundation Graduate Research Fellowship.  This work was also supported
by NSF grant PHY90-08502.

\appendix
\section*{Calculation of $\chi_\ap\circ\rho_{\lowercase{fi}}\circ\chi_\a$}

     To calculate the elements of the decoherence functional for
Sec.~\ref{antisymres}, we first expand our
notational convention for the branches to include
$\chi_{-1}\equiv\chi_l$
and $\chi_{+1}\equiv\chi_r$ so that we can write
$\chi_\lambda(x)=\Theta(\lambda
x_n)\chi(x)$, where $\lambda ^2=1$.  The  nonvanishing elements
of the decoherence functional are now 
\begin{equation}
D(\lo,\lt)=\chi_\lt\circ\rho_{fi}\circ\chi_\lo,
\end{equation}
where the inner product is on the surface $\sigma ''$.

     If $\psi$ is a solution to the Klein-Gordon equation, we
know that $(\rho_+-\rho_-)\circ\psi=\psi_++\psi_-=\psi$.  This
will not be true for $\chi_\lambda$ because it is {\em not} a
solution.  However, for the purposes of the Klein-Gordon inner
product on the surface $\sigma ''$, we only need the value and
normal derivative of each function on $\sigma ''$.  We can thus
replace $\chi_\lambda$ by $X_\lambda$, a Klein-Gordon
wave function\footnote{It is straightforward to show that such a
wave function exists, and is uniquely given by
$X_\lambda=(\rho_+-\rho_-)\circ\chi_\lambda$.} which matches
$\chi_\lambda$ and its normal derivative on $\sigma ''$.  This
gives us
\begin{eqnarray}
\chi_\lt\circ\chi_\lo &=& \chi_\lt\circ X_\lo =
\chi_\lt\circ(\rho_+-\rho_-)\circ X_\lo \nonumber\\
&=&\chi_\lt\circ(\rho_+-\rho_-)\circ\chi_\lo
\end{eqnarray}
     We thus have
\begin{eqnarray}
D(\lo,\lt)&=&\chi_\lt\circ(\rho_++\rho_-)\circ\chi_\lo \nonumber\\
&=&\chi_\lt\circ\chi_\lo + 2\chi_\lt\circ\rho_-\circ\chi_\lo.
\end{eqnarray}

     The first term is simple enough to calculate:
\begin{eqnarray}
&&\chi_\lt\circ\chi_\lo\nonumber\\
&&=\int_{\sigma ''}d^D\!
\Sigma''{}^\mu\Theta(\lt x''_n)
\chi^+(x'')^* i\nabboth''_\mu \Theta(\lo x''_n) \chi^+(x'').
\end{eqnarray}
again, since we can choose $\sigma ''$ to satisfy $n_\mu
d^D\!\Sigma''{}^\mu=0$,
we can move the step functions through the derivative to get
\begin{eqnarray}
&&\chi_\lt\circ\chi_\lo\nonumber\\
&&=\int_{\sigma ''}d^D\!
\Sigma''{}^\mu \Theta(\lt x''_n) \Theta(\lo x''_n) \chi^+(x'')^*
i\nabboth''_\mu \chi^+(x'') \nonumber\\
&&=\delta_{\lo\lt}\int_{\sigma ''}d^D\! \Sigma''{}^\mu
\Theta(\lo x''_n) \chi^+(x'')^* i\nabboth''_\mu \chi^+(x'').
\end{eqnarray}
The symmetry of $\sigma''$ and antisymmetry of $\chi^+$ tell us
that $\chi_l\circ\chi_l=\chi_r\circ\chi_r$, so
\begin{equation}
\chi_\lt\circ\chi_\lo=\delta_{\lo\lt}{\chi^+\circ\chi^+\over 2}
={\delta_{\lo\lt}\over 2}.
\end{equation}

     To calculate the correction term $\chi_\lt\circ\rho_-
\circ\chi_\lo$, we first
define the Fourier transform
of $\chi^+$ by
\begin{equation}\label{Fourier}
\chi^+(x)=\int {d^D\! k\over
(2\pi)^{D/2}\sqrt{2\ok}}
e^{i\k\cdot\x}e^{-i\ok t}\tilde\chi^+({\bf k}).
\end{equation}
The inner
product of two positive energy states is expressed in terms of
the Fourier transform by:
\begin{equation}
\varphi^+\circ\psi^+=\int d^D\! k\,\varphi^+({\bf
k})^*\psi^+({\bf k})
\end{equation}
so the normalization condition (\ref{norm}) is written as
\begin{equation}
\int d^D\! k\,|\chi^+({\bf k})|^2 = 1
\end{equation}

     In a reference
frame where $n$ is has no time component, we can split the
spatial part $\bf v$ of a vector $v$ into components along $\bf
n$: ($v_n={\bf n\cdot v}=n\cdot v$) and perpendicular to $\bf n$:
(${\bf v_\perp} = {\bf v}-v_n \bf n$).  In analogy to $v_c$
defined in Sec.~\ref{images}, we define
${\bf v}_c = {\bf v}-2 v_n{\bf n} = -v_n{\bf n} + {\bf v}_\perp$.

     $\tilde\chi^+$ is determined from $\chi^+$ by
\begin{equation}
\tilde\chi^+({\bf k})=\sqrt{2\ok}e^{i\ok t}\int {d^D\!x \over (2\pi)^{D/2}}
e^{-i\k\cdot\x}\chi^+(x),
\end{equation}
so $\tilde\chi$ obeys an antisymmetry property similar to
(\ref{chi}):
\begin{equation}\label{tchi}
\tilde\chi^+({\bf k}_c)=-\tilde\chi^+({\bf k})
\end{equation}

     To proceed any further, we would like an explicit form for the
surface $\sigma''$.  The simplest would be that $\sigma''$ is a
surface of constant time $t''$.  However, that condition would not be
Lorentz invariant, as it would pick out a reference frame in which the
final surface was one of constant time.  We know from section
\ref{surfcons} that we are only restricted in the choice of $\sigma''$
by the form of the $(D-1)$-surface $\sigma''\cap\partial R(n)$.  If we
restrict our attention to choices of $\sigma''\cap\partial R(n)$ which
are flat (a suitably invariant condition), we can always work in a
reference frame in which $\sigma''$ {\em is} a surface of constant
time.  Since we construct $\sigma''$ so that $n$ lies in it, this is
consistent with the assumption that $n$ has no time component.

     Subject to the condition of $\sigma''$ being flat\footnote{Note
that if $D=1$, $\sigma''\cap\partial R(n)$ is a point, so this holds
trivially.}, then, we can work in a reference frame where it is to be
a surface of constant time, so that
\begin{equation}
\varphi\circ\psi=\left.\int d^D\! x\,\varphi({\bf
x},t)^*i\parboth_t\psi({\bf x},t)\right|_{t=t''}.
\end{equation}
\widetext
The definition (\ref{rhom}) of $\rho_-$ means that 
\begin{equation}
\chi_\lt\circ\rho_-\circ\chi_\lo=\int{d^D\! p\over(2\pi)^D 2\op}
\left(e^{i\p\cdot\x}e^{i\op t}\circ\chi_\lo\right)
\left(e^{i\p\cdot\x}e^{i\op t}\circ\chi_\lt\right)^*.
\end{equation}
Now,
\begin{equation}\label{expchi}
e^{i\p\cdot\x}e^{i\op t}\circ\chi_\lambda=
\int {d^D\! k\over (2\pi)^{D/2}\sqrt{2\ok}}\tilde\chi^+({\bf k})
(\ok-\op)e^{-i(\ok+\op)t''} \int d^D\! x\Theta(\lambda x_n)
e^{i(\k-\p)\cdot\x};
\end{equation}
the integral over ${\bf x_\perp}$ gives
$(2\pi)^{D-1}\delta^{D-1}({\bf k_\perp-p_\perp})$, and the
integral over $x_n$ gives
\begin{eqnarray}
\lefteqn{\int_{-\infty}^\infty dx_n\Theta(\lambda
x_n)e^{i(k_n-p_n)x_n}=
\int_{-\lambda\infty}^{\lambda\infty} \lambda\,
dx_n\Theta(x_n)e^{i\lambda(k_n-p_n)x_n}} \nonumber\\
 & & =\int_0^\infty dx_n e^{i\lambda(k_n-p_n)x_n}=
{i \over \lambda(k_n-p_n)+i\varepsilon}=
{i\lambda \over k_n-p_n}+\pi\delta(k_n-p_n).
\end{eqnarray}
Substituting into (\ref{expchi}) gives
\begin{equation}
e^{i\p\cdot\x}e^{i\op t}\circ\chi_\lambda=
i\lambda\int {dk_n\over\sqrt{2\ok}}(2\pi)^{D/2-1}
\tilde\chi^+({\bf k})\left({\ok-\op \over
k_n-p_n}\right)e^{-i(\ok+\op)t''}
\end{equation}
with ${\bf k_\perp=p_\perp}$.  We thus have
\begin{eqnarray}
\chi_\lt\circ\rho_-\circ\chi_\lo &=&
\lo\lt\int{dk_{1n}dk_{2n}d^{D-1}p_\perp\over
2\sqrt{\omega_1\omega_2}(2\pi)^2}
\tilde\chi^+({\bf k}_2)^*\tilde\chi^+({\bf
k}_1)e^{-i(\omega_1-\omega_2)t''}
\nonumber\\
&&\times
\int_{-\infty}^\infty{dp_n\over 2\op}\left({\op-\omega_1 \over
p_n-k_{1n}}\right)\left({\op-\omega_2 \over p_n-k_{2n}}\right),
\label{star}
\end{eqnarray}
\narrowtext
where ${\bf k}_{1\perp}={\bf k}_{2\perp}={\bf p}_\perp$ so that
$\omega_1=\sqrt{k_{1n}^2+\omega_\perp^2}$ and
$\omega_2=\sqrt{k_{2n}^2+\omega_\perp^2}$ where
$\omega_\perp=\sqrt{{\bf p}_\perp^2+m^2}$.
The integrand of the $p_n$ integral,
\begin{equation}
f(p_n)={1\over 2\op}\left({\op-\omega_1 \over
p_n-k_{1n}}\right)\left({\op-\omega_2 \over p_n-k_{2n}}\right)
\end{equation}
is analytic (since the singularities at $p_n=k_{1n}$ and $p_n=k_{2n}$
are removable) except for branch points when $\op=0$, namely at
$p_n=i\omega_\perp$ and $p_n=-i\omega_\perp$.  We can thus deform the
integration contour to the one shown in Fig.~\ref{contour}.  The
contributions from the quarter-circle arcs cancel, and the
contributions from the branch cut give
\begin{equation}\label{fintbranch}
\int_{-\infty}^\infty f(p_n) dp_n=
\int_{\omega_\perp}^\infty{d\kappa\over\sqrt{\kappa^2-\omega_\perp^2}}
{\omega_1\omega_2+\omega_\perp^2-\kappa^2 \over (i\kappa-k_{1n})
(i\kappa-k_{2n})}.
\end{equation}
With the substitution $\kappa=\omega_\perp\sec{\theta}$, this
becomes
\begin{equation}
\int_0^{\pi/2}{\cos{\theta}
(\omega_1\omega_2+\omega_\perp^2-\omega_\perp^2\sec^2{\theta})
\over (i\omega_\perp-k_{1n}\cos{\theta})
(i\omega_\perp-k_{2n}\cos{\theta})} d\theta,
\end{equation}
which can be evaluated to give
\begin{eqnarray}
&&\int_{-\infty}^\infty f(p_n) dp_n\nonumber\\
&&=\int_0^{\pi/2}\sec{\theta}d\theta
+ {\omega_1+\omega_2\over k_{1n}-k_{2n}}
\ln\left({\omega_1-k_{1n}\over\omega_2-k_{2n}}\right).
\end{eqnarray}
The first term is a constant, and is thus even in $k_{1n}$.  The
rest of (\ref{star}) is odd in $k_{1n}$ because of (\ref{tchi})
so the constant term gives no contribution to
$\chi_\lt\circ\rho_-\circ\chi_\lo$, and
\widetext
\begin{equation}
\chi_\lt\circ\rho_-\circ\chi_\lo=
\lo\lt\int{dk_{1n}dk_{2n}d^{D-1}k_\perp\over (2\pi)^2}
{\omega_1+\omega_2\over 2\sqrt{\omega_1\omega_2}}
\tilde\chi^+({\bf k}_2)^*\tilde\chi^+({\bf
k}_1){e^{-i(\omega_1-\omega_2)t''}
\over k_{1n}-k_{2n}}
\ln\left({\omega_1-k_{1n}\over\omega_2-k_{2n}}\right).
\end{equation}
\narrowtext
This gives us (\ref{result}).

\begin{figure}
\caption{The contour on which the integral in (\protect\ref{star})
is calculated to give (\protect\ref{fintbranch}).  The radius $R$ of
the quarter-circle arcs is to be taken to infinity.}
\label{contour}
\end{figure}

\end{document}